\documentclass[
floatfix,
aps,
prl,
amsmath,
twocolumn,
superscriptaddress,
]{revtex4}

\usepackage{epsfig}
\usepackage{epsf}
\usepackage{array}

\newcommand{\mls}{\delta_1} 
\newcommand{\req}[1]{Eq.~(\ref{#1})}
\newcommand{\res}[1]{Eqs.~(\ref{#1})}
\newcommand{\kB}{k_{\rm B}}

\begin{document}

\title{Photovoltaic and Rectification Currents in
Quantum Dots}

\author{M. G. Vavilov}
\altaffiliation[Present address: ]
{Department of Applied Physics, Yale University, New Haven, CT  06520}
\affiliation{Center for Materials Sciences and Engineering,
  Massachusetts Institute of Technology, Cambridge, MA 02139}
\author{L. DiCarlo} \affiliation{Department of Physics,
Harvard University, Cambridge, MA 02138}
\author{C. M. Marcus}
\affiliation{Department of Physics, Harvard University, Cambridge, MA 02138}

\begin{abstract} We investigate
theoretically and experimentally the statistical properties of dc
current through an open quantum dot
subject to ac excitation of a shape-defining gate. The symmetries of rectification
current
and photovoltaic current with respect to applied magnetic field are examined. Theory
and experiment are found to be in good agreement throughout a broad
range of frequency and ac power, ranging from
adiabatic to nonadiabatic regimes.

\end{abstract}
\date{October 1, 2004}
\maketitle

Transport in mesoscopic systems subject to
time-varying fields combines elements of
non-equilibrium physics
and quantum chaos.
This combination extends
the scope of mesoscopic physics and is likely to be important in
quantum information processing, where fast gating and
quantum coherence are both required. Of particular importance is the ability to control
external fields applied to the mesoscopic system and
to distinguish effects of these fields on quantum dynamics
of the system. For example, two distinct contributions to
direct current through an open quantum dot due to an
oscillating perturbation have been identified \cite{Brouwer98,BrouwerR} and
observed experimentally in Ref.~\cite{DCMH}.

In this Letter, we investigate the statistical properties of dc currents resulting
from
an applied ac electric field over a wide range of excitation frequencies,
paying particular attention to
the presence or absence of symmetry with
respect to magnetic field in various regimes. Theoretical
analysis is based on recently developed time-dependent random matrix
theory~\cite{VavDeph,VavPV}. Experiments use a
gate-defined GaAs quantum dot subject to ac excitation of a gate at MHz to
GHz frequencies. At low excitation frequencies $\omega\ll \tau_{\rm d}^{-1}$
($\tau_{\rm d}$ is
the electron dwell time in the dot) the present theoretical results
are consistent with those obtained by adiabatic
approximations~\cite{Brouwer98,SAA}.
However, the analysis is applicable over a wider range of
frequencies $\omega\lesssim E_{\rm T}$, where
$E_{\rm T}=\hbar / \tau_{\rm cross}$ is the Thouless energy and
$\tau_{\rm cross}$ is the electron crossing time of the dot.
At higher frequencies $\omega\gtrsim E_{\rm T}$, the system
may be studied by methods developed for bulk
conductors~\cite{AAKL,Kravtsov01}.

Three distinct contributions to dc current through the dot can be
identified, resulting from: \emph{i})  an applied dc bias;
\emph{ii}) an ac bias at the excitation frequency (i.e.,
rectification effects~\cite{BrouwerR}); \emph{iii}) photovoltaic
effects~\cite{Falko89,VavPV}. We restrict our attention to
one-parameter excitation, noting that while in the adiabatic
regime one- and two-parameter excitations affect the system
differently, beyond the adiabatic regime,  $\omega\gtrsim
\tau_{\rm d}^{-1}$, the differences disappear~\cite{VavPV}.

\begin{figure}[htbp]
\includegraphics[width=3.05in]{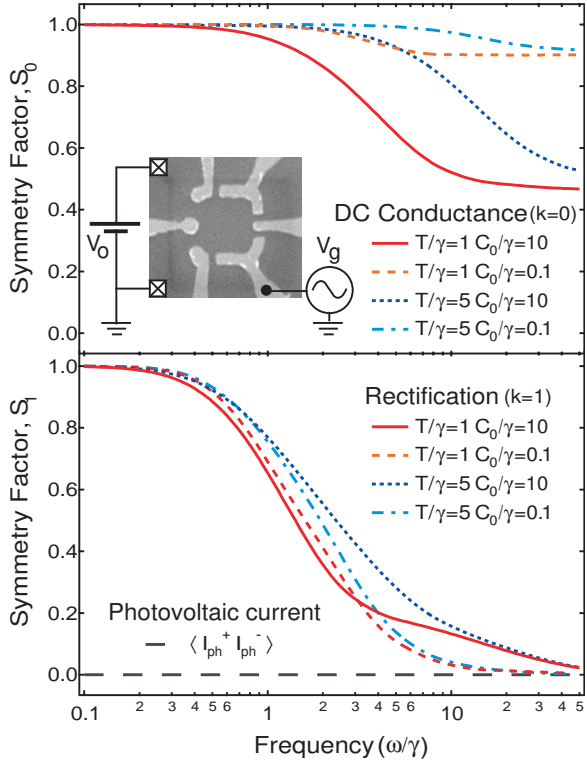}
\caption{\footnotesize{Symmetry factor $S_k=\langle
\overline{\delta G_k^{+\Phi}} \
\overline{\delta G_k^{-\Phi}}\rangle /
\langle (\overline{\delta G_k^\Phi})^2 \rangle$
 as a function of frequency $\omega$ for $k=0$ (upper panel)
 and $k=1$ (lower panel) at two values of temperature $T$ and
 power $C_0$ of the ac excitation.
 Inset: Micrograph of device and schematic picture of
 applied voltages.}
}
\label{Figure1}
\end{figure}

The Hamiltonian of electrons in the dot
in the presence of a magnetic flux $\Phi$
is represented by a Hermitian $M\times M$ matrix
$\hat H(t)=\hat {\cal H}_\Phi+\hat {\cal V}\cos\omega t$,
with the time independent part $\hat {\cal H}_\Phi$ being a random
realization of a matrix from a Gaussian unitary ensemble
with the mean level spacing $\mls$,
and $\hat {\cal V}$ being a matrix from a Gaussian orthogonal ensemble
characterized by the strength $C_0=\pi{\rm Tr}\hat {\cal V}^2/M^2\mls$
and $M\mls \sim E_{\rm T}$ \cite{floq}.
The parameter $C_0$ determines the energy displacement of an electron
state due to the applied perturbation $\hat {\cal V}$.
The contact between the left (right) lead and the dot
contains $N_{\rm l}$ ($N_{\rm r}$) open channels,
we enumerate channels, $\alpha$, in the left ($\alpha=1\dots N_{\rm l}$)
and the right ($\alpha= N_{\rm l}+1\dots N_{\rm ch}$) contacts,
$N_{\rm ch}=N_{\rm l}+N_{\rm r}$. The corresponding
experimental setup is shown in Fig.~\ref{Figure1}.

The dc current $\bar I^\Phi$ through
the dot is determined by
the scattering matrix $[{\cal S}_\Phi(t,t')]_{\alpha\beta}$, see Ref.~\cite{VavPV}:
\begin{equation}
\begin{split}
\bar I^{\Phi} & = \frac{e\omega}{2\pi} \int\limits_{0}^{2\pi/\omega}
dt\int\limits_{-\infty }^{+\infty }dt_{1}dt_{2}
\\ &\times
{\rm Tr}\left\{ \hat{f}(t_{1},t_{2}) \left[ \hat{{\cal
S}}^{\dagger }_{\Phi}(t_{2},t)
\hat{\Lambda}\hat{{\cal S}}_{\Phi}(t,t_{1}) -
\hat{\Lambda} \delta_{t,t_1}\delta_{t,t_2}\right]\right\}.
\end{split}
\label{Idcgeneral}
\end{equation}
Here $\delta_{t,t'}=\delta(t-t')$ and
\begin{equation}
f_{\alpha\beta}(t,t')=
\frac{\kB T}{\hbar}\frac{\delta_{\alpha\beta}\exp\left(
i\frac{e}{\hbar}\int_{t'}^t V_\alpha(\tau)d\tau\right)}
{\sinh (\pi \kB T(t-t')/\hbar)}
\label{fdistr}
\end{equation}
is the distribution function of electrons
in channel $\alpha$ at temperature $T$
and voltage $V_\alpha(t)$.
At sufficiently low frequencies $\omega\ll E_c/\hbar$
($E_c$ is the dot charging energy)
$V_\alpha(t)$ is simply related to the bias $V(t)$
across the dot: $V_\alpha(t)=\Lambda_{\alpha\alpha}V(t)$.
Elements of the diagonal matrix $\hat \Lambda$ are
$\Lambda_{\alpha\alpha}=N_{\rm r}/N_{\rm ch}$ for
$1\leq \alpha \leq N_{\rm l}$, and
$\Lambda_{\alpha\alpha}=-N_{\rm l}/N_{\rm ch}$ for
$N_{\rm l} < \alpha \leq N_{\rm ch}$.

We consider the bias $V(t)$ across the dot in the form
$V(t)=V_0+V_{\omega}\cos(\omega t+\varphi_1)$.
The dc current through the dot to first order in
dc bias $V_0$ and ac bias $V_\omega$
is \cite{nonlin}
\begin{equation}
\bar I^{\Phi}=\bar I^{\Phi}_{\rm ph}+\bar I_1^{\Phi} + \bar g^{\Phi}_0V_0,
\quad
\bar I_1^{\Phi}= \bar g^{\Phi}_1V_\omega,
\label{Idc}
\end{equation}
where the first term represents the photovoltaic current
$\bar I^{\Phi}_{\rm ph}=\bar I^{\Phi}(V_{\alpha}\equiv 0)$,
see \res{Idcgeneral} and (\ref{fdistr}).
The second and third terms in \req{Idc} represent the contributions to
the current due to dc bias $V_0$ and ac bias $V_\omega$, respectively:
\begin{equation}
\bar g_{0}^{\Phi} = \frac{e^2}{\pi \hbar}
\left[G_0 -
\overline{\delta G_{0}^{\Phi}(t)}
\right],\quad
\bar g_1^{\Phi} = -\frac{e^2}{\pi \hbar}\overline{\delta G_{1}^{\Phi}(t)},
\label{Ibias01}
\end{equation}
where $G_0=N_{\rm l}N_{\rm r}/N_{\rm ch}$ is the classical conductance and
$\overline{\delta G_{k}(t)}=\int_0^{2\pi/\omega}\delta G_{k}(t)\omega dt/2\pi$
stands for time averaging of
the ``instantaneous conductance'' ($k=0,1$)
\begin{eqnarray}
\delta G_{k}^{\Phi}(t)  & = &
\label{Ibias}
\int
 d\tau F_{k}(\tau)\int d\theta \cos(k(\omega\theta+\varphi_k))
\label{Ibias2}
\\
&  \times& \!
{\rm Tr}\left\{
\hat{\Lambda}
\hat{{\cal S}}_{\Phi}^{\dagger }\left(\theta-\frac{\tau}{2},t\right)
\hat{\Lambda}
\hat{{\cal S}}_{\Phi}\left(t,\theta+\frac{\tau}{2}\right)
\right\},
\nonumber
\\
F_{0}(\tau) &\!\! = &\!\! \! \frac{\pi \kB T\tau/\hbar }{\sinh(\pi\kB T\tau/\hbar)},
\
F_{1}(\tau) =  \frac{2\pi \kB T\sin(\omega\tau/2)}
{\hbar\omega\sinh(\pi\kB T\tau/\hbar)}
.
\nonumber
\end{eqnarray}
We observe that
$
\overline{\delta G_1(t)}=\overline{\delta G_0(t)\cos(\omega t+\varphi_1)}
$
in the adiabatic limit
$\hbar \omega\ll {\rm max}\{N_{\rm ch}\mls, \kB T\}$
considered in Refs.~\cite{BrouwerR,Moskalets03}.


Below we study the variance of the photovoltaic current $\bar I_{\rm ph}^\Phi$
and the conductances $\bar g_0^{\Phi}$ and $\bar g_1^{\Phi}$
with respect to random realizations of the Hamiltonian $\hat {\cal H}_{\Phi}$.
Following Refs.~\cite{VavPV,VavOns}, we find
in the limit $N_{\rm ch}\gg 1$ and at magnetic fields
$\Phi \gg \Phi_0\sqrt{N_{\rm ch}/M}$ destroying the weak
localization ($\Phi_0=hc/e$)
\begin{eqnarray}
\big{\langle}\!\! \left(\bar I_{\rm ph}^{\Phi}\right) ^{\! 2}\! \big{\rangle}
=\frac{G_0 e^2 \omega^4 C_0\mls}{2\pi^3\hbar^2}  \!\!
\!\!\!  \int\limits_0^{2\pi/\omega} \!\!\! \frac{dt dt'}{4\pi^2}
\!\! \int\limits_{0}^{\infty} \!\! d\tau \!\!\!
\int\limits_{\tau/2}^{\infty}\!\!\! d\theta\!K^+
B^{\rm ph}_{t-\theta,t'-\theta;\tau},&&
\label{Rph}
\\
\big{\langle}\!
\overline{\delta G_k^{+\Phi}}\ \overline{\delta G_k^{\pm\Phi}}
\! \big{\rangle}
= \frac{G_0^2\omega^2\mls^2}{2\pi^2\hbar^2}\!\!\!\!
\int\limits_0^{2\pi/\omega} \!\! \frac{dt dt'}{4\pi^2} \!\!\!
\int\limits_{0}^{\infty} \!\! d\tau \!\!
\int\limits_{\tau/2}^{\infty}\!\! d\theta\,
K^\pm
B^{(k)}_{t-\theta,t'-\theta;\tau}&&
\label{RbiasAnti}
\label{Rbias}
\end{eqnarray}
and
$\langle \bar I^{+\Phi}_{\rm ph} \bar I^{-\Phi}_{\rm ph}
\rangle=0$, see Ref.~\cite{SAA}.
In \res{Rph} and (\ref{RbiasAnti}) the angle brackets $\langle\dots\rangle$
stand for the averaging with respect to realizations of $\hat {\cal H}_\Phi$.
Functions $B^{(k)}_{t,t';\tau}$ and $B^{\rm ph}_{t,t';\tau}$ describe
the distribution function~\cite{footnoteB} of electrons in the dot in the
presence of time-dependent electric fields
and kernels $K^{\pm}\equiv K^{\pm}_{t,t';\theta,\tau}$
describe the evolution of electron states~\cite{footnoteK} in these
fields. Both functions
$B^{(k)}_{t,t';\tau}$ and $B^{\rm ph}_{t,t';\tau}$ and the kernels
$K^{\pm}_{t,t';\theta,\tau}$ contain the diffuson
$
{\cal D}(t_1,t_2,\tau) = \exp\left(
-\int_{t_2}^{t_1}\Gamma(\tau,t)dt \right)
$
or the Cooperon
$
{\cal C}(\tau_1,\tau_2,t) = \exp\left(
-\frac{1}{2}\int_{\tau_2}^{\tau_1}\Gamma(\tau,t)d\tau
\right)
$.
Here,
$
\Gamma(\tau,t)=\gamma_{\rm e}+\gamma_\varphi
+4C_0\sin^2\omega t\sin^2(\omega\tau/2)
$, where
$\gamma_{\rm e}=\mls N_{\rm ch}/2\pi$ is the electron escape rate
and $\gamma_\varphi$ is the electron phase relaxation rate due
to inelastic processes.

\begin{figure}[t]
\includegraphics[width=3.25in]{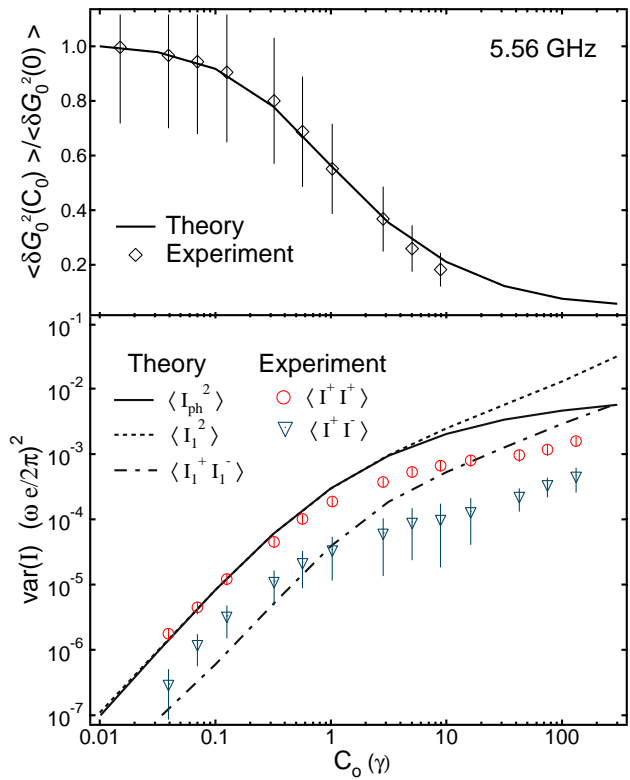}
\caption{\footnotesize{Upper panel: Variance of the conductance as
a function of the ac excitation power $P=P_0C_0/\gamma_{\rm e}$ at
$\omega/2\pi=5.56\ $GHz ($\hbar\omega=7.2\gamma_{\rm e}$) and the
theoretical result of \req{Rbias} with $k=0$. We use $P_0=9\times
10^{-8}\ $W and $\kB T=5.4\gamma_{\rm e}$. Lower panel: Symmetric
($\bigcirc$) and antisymmetric ($\bigtriangledown$) current
correlators as a function of ac excitation strength $C_0$. Solid
line shows variance of the photovoltaic current \req{Rph} with
parameters fixed by the fit in the upper panel. The dashed and
dotted lines show the symmetric and antisymmetric correlators of
the rectification current \req{Rrect} with
$\alpha_\omega=0.45\hbar \omega/e$ and $\varphi_1=0$. }}
\label{Figure3}
\end{figure}

In the experiment, the ac bias $V_\omega$
results from capacitive coupling between the leads and the gate on which
the ac voltage is applied (see the inset in Fig. 1).
Therefore $V_\omega$ is proportional to the amplitude of the
ac voltage at the gate. Assuming that $C_0$
is linear in the applied power to the gates, we write
$V_\omega=\alpha_{\omega}\sqrt{C_0/\gamma_{\rm e}}$.
The coefficient $\alpha_\omega$ has units of voltage and
is independent of realizations of the quantum
dot (we disregard fluctuations of $\hat {\cal V}$
over different realizations of $\hat {\cal H}_\Phi$).
Therefore, the correlation functions of the rectification
current $\bar I^{\Phi}_1=\bar g_1^{\Phi}V_\omega$ are determined by
the correlators of $\overline{\delta G_1^{\Phi}}$:
\begin{equation}
\Big{\langle}
\bar I^{+\Phi}_1\bar I^{\pm\Phi}_1
\Big{\rangle}
=\frac{e^4}{\pi^2\hbar^2} \alpha_\omega^2
\ \frac{C_0}{\gamma_{\rm e}}\ \Big{\langle}
\overline{\delta G^{+\Phi}_1}\
\overline{\delta G^{\pm\Phi}_1}
\Big{\rangle}.
\label{Rrect}
\end{equation}
We also notice that in the limit $N_{\rm ch}\gg 1$ the
correlation function of the photovoltaic current $\bar I_{\rm ph}^\Phi$
and the rectification current $\bar I_1^\Phi$ vanishes~\cite{SAA}.

First we use \res{Rbias} and (\ref{Rrect}) to analyze
the magnetic field symmetry of
the rectification current $\bar I_{1}^\Phi$.
Although in the adiabatic limit $\omega \ll \gamma_{\rm e}/\hbar $
the rectification current
$\bar I_1^\Phi$ is symmetric with respect to magnetic field inversion
($\Phi\to -\Phi$), at higher frequencies
$\omega\gtrsim \gamma_{\rm e}/\hbar$ the symmetry of $\bar I_1^\Phi$
is suppressed. Indeed,
the magnetic field symmetry is related to the time inversion
symmetry. For a harmonic field at frequency $\omega$,
the time-inversion symmetry holds only on time scales much smaller than
$1/\omega$. Transport through the system is determined
by times of the order of $\hbar/\gamma_{\rm e}$ and consequently
the magnetic field symmetry of the rectification current breaks
if $\hbar \omega \gtrsim \gamma_{\rm e}$. We plot the ratio
$S_1=\langle \overline{\delta G_1^{+\Phi}} \
\overline{\delta G_1^{-\Phi}}\rangle /
\langle (\overline{\delta G_1^\Phi})^2 \rangle$
as a function of $\hbar \omega/\gamma_{\rm e}$ in Fig.~\ref{Figure1}.
$S_1=1$ represents the symmetric current
$\bar I_1^{\Phi}\propto \overline{\delta G_1^{\Phi}}$
with respect to magnetic field inversion.
In the adiabatic regime this symmetry originates from
the Onsager symmetry~\cite{Onsager} of the dc conductivity,
see \req{Ibias} and Refs.~\cite{BrouwerR,DCMH}.
As the frequency increases, $S_1$ vanishes, signalling the
suppression of the magnetic field symmetry.
Therefore, the absence of magnetic field symmetry no longer serves as a
distinct feature of the photovoltaic current $\bar I_{\rm ph}^\Phi$,
which allows one to distinguish $\bar I_{\rm ph}^\Phi$
and the rectification current~$\bar I_1^\Phi$.

We notice that the magnetic field symmetry of the dc conductance
$\overline{\delta G_0^{\Phi}}$
is more sturdy than the symmetry of the rectification current,
see Fig.~\ref{Figure1}. Particularly, at
temperatures $\kB T\gtrsim \gamma_{\rm e}$,
dc conductance $\overline{\delta G_0^{\Phi}}$
is nearly symmetric at
frequencies $\hbar \omega\lesssim \kB T$, since the dc
correlation function is determined by processes
on a time scale $\hbar/\kB T$. The symmetry is not fully suppressed
even at $\omega\gg \kB T/\hbar$; the suppression depends on
$C_0/\gamma_{\rm e}$.

We apply \res{Rph} - (\ref{Rrect}) to the analysis of
the experiment~\cite{DCMH}.
The quantum dot used in the experiment has an area
$A\approx 0.7$~$\mu$m$^2$. Relevant energy scales are
the Thouless energy
$E_{\rm T}\approx 160$~$\mu$eV
and the mean level spacing
$\mls=\pi\gamma_{\rm e}\approx 10$~$\mu$eV.
The measurements were performed
at the base electron temperature $T\approx 200$~mK
($\kB T=5.4\gamma_{\rm e}$).
From the size of conductance fluctuations without ac fluctuation
on the gate, we estimate the dephasing rate
$\gamma_{\varphi}\approx 0.2 \gamma_{\rm e}$
(see~\cite{HFPMDH} for details) and thus disregard it in our
quantitative analysis.

In the upper panel of Fig.~\ref{Figure3} we show the variance of
the conductance as a function of the incident power $P$ at
$\omega/2\pi=5.56$~GHz ($\hbar \omega/\gamma_{\rm e}\approx 7.2$).
We also plot the variance of the conductance calculated from
\req{Rbias} ($k=0$) at temperature $T=5.4\gamma_{\rm e}/\kB$.
Assuming that the ratio $C_0/\gamma_{\rm e}$ is proportional to
the power $P$ of the ac excitation applied to the gate,
i.e. $C_0/\gamma_{\rm e}=P/P_0$, we rescale $P$ to
obtain the best fit of the experimental points by the curve of
\req{Rbias}. We find $P_0=9\times 10^{-8}$~W.

\begin{figure}[t]
\includegraphics[width=3.25in]{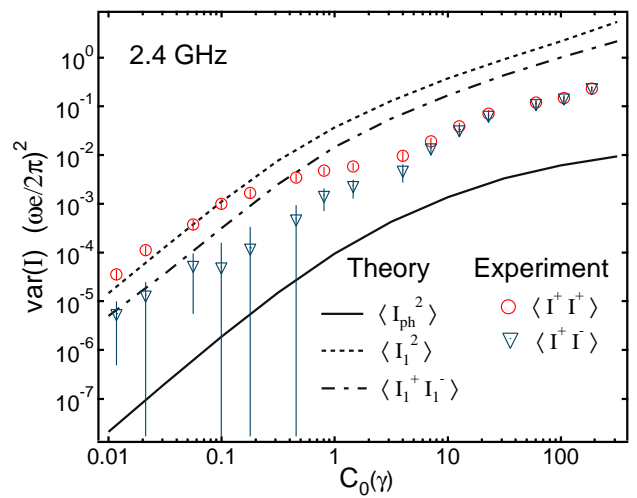}
\caption{\footnotesize{Symmetric ($\bigcirc$) and antisymmetric
($\bigtriangledown$) current correlators at $\omega/2\pi =2.4\
$GHz as a function of power $P=P_0C_0/\gamma_{\rm e}$ with
$P_0=2.5\times 10^{-7}\ $W. Solid line shows variance of the
photovoltaic current \req{Rph} at temperature $T=5.4\gamma_{\rm
e}/\kB$ and frequency $\omega=3.1\gamma_{\rm e}/\hbar$. The dashed
and dotted lines show the symmetric and antisymmetric correlators
of the rectification current \req{Rrect} with $\alpha_\omega=4.7
\hbar \omega/e$ and $\varphi_1=0$.}} \label{Figure4}
\end{figure}

In the lower panel of Fig.~\ref{Figure3} we show the correlators
$\langle \bar I^{+\Phi}\bar I^{\pm\Phi}\rangle$
of the measured current. Although the traces of the
magnetic field sweeps look quite asymmetric for the measured
current, the antisymmetric correlator
$\langle \bar I^{+\Phi}\bar I^{-\Phi}\rangle$ is not significantly
smaller than its symmetric counterpart. We notice however, that if
the averaging is performed over $n$ realizations, the measured
correlator $\langle \bar I^{+\Phi}\bar I^{-\Phi}\rangle$
can be estimated as
$\langle \bar I^{+\Phi}\bar I^{+\Phi}\rangle/\sqrt{n}$
($n\sim 50$ in the experiment).

We plot the variance of the photovoltaic current,
\req{Rph}, as a function of $C_0/\gamma_{\rm e}$ for
$\kB T=5.4\gamma_{\rm e}$ and $\hbar \omega=7.2\gamma_{\rm e}$
(to facilitate numerics, we used the approximation
$\hbar \omega\gg \gamma_{\rm e}$).
We emphasize that the horizontal shift between the data
points and curve is fixed by the fit in the upper panel for the dc
conductance and there is no fitting parameters for the variance of
the  current (along vertical axis).
At $C_0\lesssim \gamma_{\rm e}$,
the variance of the measured current changes
quadratically in $C_0/\gamma_{\rm e}$, consistent with
quadratic dependence on $C_0$ of the theoretical curves for
$\langle \bar I_{\rm ph}^2\rangle$ and therefore
our assumption that $C_0$ is proportional to the
power of the ac excitation is justified.
At $C_0\gtrsim \gamma_{\rm e}$ the variance of the measured current
starts saturating. This saturation is expected for large power
asymptote of the photovoltaic current
due to spreading of the distribution function
of electrons in the dot~\cite{VavPV}.
Some deviation  between the experimental points and the theoretical curve
is expected due to the approximation $N_{\rm ch}\gg 1$ used for
derivation of \req{Rph} ($N_{\rm ch}=2$ in the experiment).

For illustration, we also plot the correlation functions
of the rectification current, using \req{Rrect} for $\varphi_1=0$
and $\alpha_\omega = 0.45 \hbar \omega/ e$.
For the rectification current, the saturation at large power
is not expected: according to Fig.~\ref{Figure3},
$\langle \bar I_{1}^{+\Phi} \bar I_{1}^{\pm\Phi}\rangle\propto
(C_0/\gamma_{\rm e})^{a}$ with $a\approx 0.6$.

We similarly discuss the data for $\omega/2\pi=2.4$~GHz.
Performing the fit of the experimental values of the conductance
fluctuations and the result of \req{Rbias} with $k=0$ and
temperature $T=5.4\gamma_{\rm e}/\kB$, we find the relation between
the strength of the perturbation $C_0$ and
the power  $P=P_0C_0/\gamma_{\rm e}$
with $P_0=2.5\times 10^{-7}$~W.

In Fig.~\ref{Figure4} we show the
symmetric and antisymmetric current correlators for $2.4$~GHz.
For comparison we plot by a solid line the
variance of the photovoltaic current $\bar I_{\rm ph}^\Phi$, calculated from
\req{Rph} at $\omega=3.1\gamma_{\rm e}$. We observe that the
fluctuations of the measured current significantly exceed
(by a factor $\sim 100$) the expected magnitude
for the photovoltaic current, and therefore are likely due to the
rectification of the bias across the dot. The low power data can
be fitted by \req{Rrect} with
$\alpha_\omega=4.7\hbar \omega/e$.

The above choice for
$\alpha_\omega\gtrsim \{\hbar\omega,\kB T\}/e$ limits the
applicability of the linear expansion \req{Idc} to small powers of
the ac excitation $C_0$, such that
$ C_0/\gamma_{\rm e}\lesssim (\hbar\omega/e \alpha_\omega)^2$.
The higher order corrections in the bias $V_\omega$
do not restore magnetic field symmetry, which is in apparent
contradiction to the observed symmetry of the measured current at
larger powers (at $C_0/\gamma_{\rm e}\gtrsim 1$ in
Fig.~\ref{Figure4}). We attribute the restoration of magnetic field
symmetry to dephasing due to dot heating by the dissipative current.
Increasing the power $P$ at fixed $\omega$
drives the system into the adiabatic regime since the heating
makes the ratio
$\hbar\omega/(\gamma_{\rm e}+\gamma_\varphi)$ decrease.
As shown already in Fig.~\ref{Figure1}, the rectification
current is symmetric in the adiabatic regime.
The assumption that $\gamma_\varphi$
increases as power $P$ increases
is consistent with the
observed change of the correlation field for the current
fluctuations, see Fig.~4 in Ref.~\cite{DCMH}.

In summary, we studied ensemble fluctuations of dc current through
an open
quantum dot subject to oscillating perturbation. We showed that as
frequency of the perturbation increases, magnetic field symmetry
of the current disappears, regardless of the mechanism of the
current generation. We demonstrated that
the power behavior of the
current fluctuations is an important tool to distinguish
effects of an ac excitation on dc current.

We thank I. Aleiner, P. Brouwer, and V. Falko for useful
discussions, and M. Hanson and A.~C.~Gossard at UC Santa Barbara for
high-quality heterostructure materials used in
the experiments. The work was supported by NSF grants DMR
02-13282 and DMR  0072777, and by AFOSR grant F49620-01-1-0475.


\begin{thebibliography}{17}
\expandafter\ifx\csname natexlab\endcsname\relax\def\natexlab#1{#1}\fi
\expandafter\ifx\csname bibnamefont\endcsname\relax
  \def\bibnamefont#1{#1}\fi
\expandafter\ifx\csname bibfnamefont\endcsname\relax
  \def\bibfnamefont#1{#1}\fi
\expandafter\ifx\csname citenamefont\endcsname\relax
  \def\citenamefont#1{#1}\fi
\expandafter\ifx\csname url\endcsname\relax
  \def\url#1{\texttt{#1}}\fi
\expandafter\ifx\csname urlprefix\endcsname\relax\def\urlprefix{URL }\fi
\providecommand{\bibinfo}[2]{#2}
\providecommand{\eprint}[2][]{\url{#2}}

\bibitem[{\citenamefont{Brouwer}(1998)}]{Brouwer98}
\bibinfo{author}{\bibfnamefont{P.~W.} \bibnamefont{Brouwer}},
  \bibinfo{journal}{Phys. Rev. B} \textbf{\bibinfo{volume}{58}},
  \bibinfo{pages}{R10135} (\bibinfo{year}{1998}).

\bibitem[{\citenamefont{Brouwer}(2001)}]{BrouwerR}
\bibinfo{author}{\bibfnamefont{P.~W.} \bibnamefont{Brouwer}},
  \bibinfo{journal}{Phys. Rev. B} \textbf{\bibinfo{volume}{63}},
  \bibinfo{pages}{121303} (\bibinfo{year}{2001}).

\bibitem[{\citenamefont{DiCarlo et~al.}(2003)\citenamefont{DiCarlo, Marcus, and
  Harris}}]{DCMH}
\bibinfo{author}{\bibfnamefont{L.}~\bibnamefont{DiCarlo}},
  \bibinfo{author}{\bibfnamefont{C.~M.} \bibnamefont{Marcus}},
  \bibnamefont{and} \bibinfo{author}{\bibfnamefont{J.~S.}
  \bibnamefont{Harris}}, \bibinfo{journal}{Phys. Rev. Lett.}
  \textbf{\bibinfo{volume}{91}}, \bibinfo{pages}{246804}
  (\bibinfo{year}{2003}).

\bibitem[{\citenamefont{Vavilov and Aleiner}(1999)}]{VavDeph}
\bibinfo{author}{\bibfnamefont{M.~G.} \bibnamefont{Vavilov}} \bibnamefont{and}
  \bibinfo{author}{\bibfnamefont{I.~L.} \bibnamefont{Aleiner}},
  \bibinfo{journal}{Phys. Rev. B} \textbf{\bibinfo{volume}{60}},
  \bibinfo{pages}{R16311} (\bibinfo{year}{1999}).

\bibitem[{\citenamefont{Vavilov et~al.}(2001)\citenamefont{Vavilov, Ambegaokar,
  and Aleiner}}]{VavPV}
\bibinfo{author}{\bibfnamefont{M.~G.} \bibnamefont{Vavilov}},
  \bibinfo{author}{\bibfnamefont{V.}~\bibnamefont{Ambegaokar}},
  \bibnamefont{and} \bibinfo{author}{\bibfnamefont{I.~L.}
  \bibnamefont{Aleiner}}, \bibinfo{journal}{Phys. Rev. B}
  \textbf{\bibinfo{volume}{63}}, \bibinfo{pages}{195313}
  (\bibinfo{year}{2001}).

\bibitem[{\citenamefont{Shutenko et~al.}(2000)\citenamefont{Shutenko, Aleiner,
  and Altshuler}}]{SAA}
\bibinfo{author}{\bibfnamefont{T.~A.} \bibnamefont{Shutenko}},
  \bibinfo{author}{\bibfnamefont{I.~L.} \bibnamefont{Aleiner}},
  \bibnamefont{and} \bibinfo{author}{\bibfnamefont{B.~L.}
  \bibnamefont{Altshuler}}, \bibinfo{journal}{Phys. Rev. B}
  \textbf{\bibinfo{volume}{61}}, \bibinfo{pages}{10366} (\bibinfo{year}{2000}).

\bibitem[{\citenamefont{Altshuler et~al.}(1982)\citenamefont{Altshuler, Aronov,
  Khmelnitskii, and Larkin}}]{AAKL}
\bibinfo{author}{\bibfnamefont{B.~L.} \bibnamefont{Altshuler}},
  \bibinfo{author}{\bibfnamefont{A.~G.} \bibnamefont{Aronov}},
  \bibinfo{author}{\bibfnamefont{D.~E.} \bibnamefont{Khmelnitskii}},
  \bibnamefont{and} \bibinfo{author}{\bibfnamefont{A.~I.}
  \bibnamefont{Larkin}}, \emph{\bibinfo{title}{Quantum Theory of Solids}}
  (\bibinfo{publisher}{Mir publisher}, \bibinfo{address}{Moscow},
  \bibinfo{year}{1982}).

\bibitem[{\citenamefont{Wang and Kravtsov}(2001)}]{Kravtsov01}
\bibinfo{author}{\bibfnamefont{X.-B.} \bibnamefont{Wang}} \bibnamefont{and}
  \bibinfo{author}{\bibfnamefont{V.~E.} \bibnamefont{Kravtsov}},
  \bibinfo{journal}{Phys. Rev. B} \textbf{\bibinfo{volume}{64}},
  \bibinfo{pages}{033313} (\bibinfo{year}{2001}).

\bibitem[{\citenamefont{Falko and Khmel'nitskii}()}]{Falko89}
\bibinfo{author}{\bibfnamefont{V.~I.} \bibnamefont{Falko}} \bibnamefont{and}
  \bibinfo{author}{\bibfnamefont{D.~E.} \bibnamefont{Khmel'nitskii}},
  \bibinfo{howpublished}{Zh. Eksp. Teor. Fiz. \textbf{95}, 328, (1989), [Sov.
  Phys. JETP \textbf{68}, 186 (1989)].}

\bibitem[{flo()}]{floq}
\bibinfo{howpublished}{Alternative methods to the time-dependent random
  matrix theory here are based on Floquet approach, see M.~Moskalets and
  M.~B\" uttiker, Phys. Rev. B {\bf 66}, 205320 (2002), or time-dependent
  scattering matrix approach, see M.L. Polianski and P.W. Brouwer,
  J. Phys. A: Math. Gen.  {\bf 36}, 3215 (2003). }

\bibitem[{non()}]{nonlin}
\bibinfo{howpublished}{The biased current produces heating, and the analysis
  beyond the linear response in the bias voltage requires consideration of heat
  relaxation in the system}.

\bibitem[{\citenamefont{Moskalets and Buttiker}(2003)}]{Moskalets03}
\bibinfo{author}{\bibfnamefont{M.}~\bibnamefont{Moskalets}} \bibnamefont{and}
  \bibinfo{author}{\bibfnamefont{M.}~\bibnamefont{B\" uttiker}},
  \bibinfo{journal}{Phys. Rev. B} \textbf{\bibinfo{volume}{69}},
  \bibinfo{pages}{205316} (\bibinfo{year}{2004}).

\bibitem[{\citenamefont{Vavilov and Aleiner}(2001)}]{VavOns}
\bibinfo{author}{\bibfnamefont{M.~G.} \bibnamefont{Vavilov}} \bibnamefont{and}
  \bibinfo{author}{\bibfnamefont{I.~L.} \bibnamefont{Aleiner}},
  \bibinfo{journal}{Phys. Rev. B} \textbf{\bibinfo{volume}{64}},
  \bibinfo{pages}{085115} (\bibinfo{year}{2001}).

\bibitem[{foo({\natexlab{a}})}]{footnoteB}
\bibinfo{howpublished}{$ B^{(k)}_{t,t';\tau} = F_{k}^2(\tau)\cos[k(\omega
  t+\varphi_k)] \cos[k(\omega t'+\varphi_k)]$ and
  $B^{\rm ph}_{t,t';\tau} =(\gamma_{\rm e}/\hbar)^2
  F_1^2(\tau)  \int\limits_0^{\infty}d\xi d\xi'
  {\cal D}(t,t-\xi,\tau){\cal D}(t',t'-\xi',\tau)\\
  \times
  \left[\!\sin \omega t \sin\omega t' +\frac{2C_0}{\gamma_{\rm
  e}}\sin^2\omega(t-\xi) \sin^2\omega(t'-\xi')\sin^2\frac{\omega\tau}{2}
  \right]
  $}.

\bibitem[{foo({\natexlab{b}})}]{footnoteK}
\bibinfo{howpublished}{$K^+_{t,t';\theta,\tau} \!\!=\!
{\cal D}(\frac{t+t'}{2},\frac{t+t'+\tau}{2}-\theta,t'-t)
{\cal D}(\frac{t+t'}{2},\frac{t+t'-\tau}{2}-\theta,t-t')\\ $ and $
K^-_{t,t';\theta,\tau}=  {\cal C}(t-t'+\theta-\frac{\tau}{2},t-t'-\theta+\frac{\tau}{2},
  \frac{t+t'-\theta}{2}+\frac{\tau}{4}) \\ \times{\cal
  C}(t'-t+\theta+\frac{\tau}{2},t'-t-\theta-\frac{\tau}{2},
  \frac{t+t'-\theta}{2}-\frac{\tau}{4}) $}.

\bibitem[{\citenamefont{Onsager}()}]{Onsager}
\bibinfo{author}{\bibfnamefont{L.}~\bibnamefont{Onsager}},
  \bibinfo{howpublished}{Phys. Rev. \textbf{38}, 2265 (1931); M. B\" uttiker,
  Phys. Rev. Lett. \textbf{57}, 1761 (1986)}.

\bibitem[{\citenamefont{Huibers }(1999)}]{HFPMDH}
\bibinfo{author}{\bibfnamefont{A.~G.} \bibnamefont{Huibers, J.A. Folk, S.R. Patel,
C.M.~Marcus, C.I.~Duru\" oz, and J.S. Harris}}, \bibinfo{journal}{Phys. Rev.
  Lett.} \textbf{\bibinfo{volume}{83}}, \bibinfo{pages}{5090}
  (\bibinfo{year}{1999}).

\end{thebibliography}

\end{document}